\documentclass[12pt]{article}
\usepackage{graphicx}
\usepackage{lscape}
\usepackage{citesort}
\usepackage{amssymb}
\usepackage{appendix}
\usepackage{multirow}

\begin{document}
\def\qq{\langle \bar q q \rangle}
\def\uu{\langle \bar u u \rangle}
\def\dd{\langle \bar d d \rangle}
\def\sp{\langle \bar s s \rangle}
\def\GG{\langle g_s^2 G^2 \rangle}
\def\Tr{\mbox{Tr}}
\def\figt#1#2#3{
        \begin{figure}
        $\left. \right.$
        \vspace*{-2cm}
        \begin{center}
        \includegraphics[width=10cm]{#1}
        \end{center}
        \vspace*{-0.2cm}
        \caption{#3}
        \label{#2}
        \end{figure}
	}
	
\def\figb#1#2#3{
        \begin{figure}
        $\left. \right.$
        \vspace*{-1cm}
        \begin{center}
        \includegraphics[width=10cm]{#1}
        \end{center}
        \vspace*{-0.2cm}
        \caption{#3}
        \label{#2}
        \end{figure}
                }

\def\ds{\displaystyle}
\def\beq{\begin{equation}}
\def\eeq{\end{equation}}
\def\bea{\begin{eqnarray}}
\def\eea{\end{eqnarray}}
\def\beeq{\begin{eqnarray}}
\def\eeeq{\end{eqnarray}}
\def\ve{\vert}
\def\vel{\left|}
\def\ver{\right|}
\def\nnb{\nonumber}
\def\ga{\left(}
\def\dr{\right)}
\def\aga{\left\{}
\def\adr{\right\}}
\def\lla{\left<}
\def\rra{\right>}
\def\rar{\rightarrow}
\def\lrar{\leftrightarrow}  
\def\nnb{\nonumber}
\def\la{\langle}
\def\ra{\rangle}
\def\ba{\begin{array}}
\def\ea{\end{array}}
\def\tr{\mbox{Tr}}
\def\ssp{{\Sigma^{*+}}}
\def\sso{{\Sigma^{*0}}}
\def\ssm{{\Sigma^{*-}}}
\def\xis0{{\Xi^{*0}}}
\def\xism{{\Xi^{*-}}}
\def\qs{\la \bar s s \ra}
\def\qu{\la \bar u u \ra}
\def\qd{\la \bar d d \ra}
\def\qq{\la \bar q q \ra}
\def\gGgG{\la g^2 G^2 \ra}
\def\q{\gamma_5 \not\!q}
\def\x{\gamma_5 \not\!x}
\def\g5{\gamma_5}
\def\sb{S_Q^{cf}}
\def\sd{S_d^{be}}
\def\su{S_u^{ad}}
\def\sbp{{S}_Q^{'cf}}
\def\sdp{{S}_d^{'be}}
\def\sup{{S}_u^{'ad}}
\def\ssp{{S}_s^{'??}}

\def\sig{\sigma_{\mu \nu} \gamma_5 p^\mu q^\nu}
\def\fo{f_0(\frac{s_0}{M^2})}
\def\ffi{f_1(\frac{s_0}{M^2})}
\def\fii{f_2(\frac{s_0}{M^2})}
\def\O{{\cal O}}
\def\sl{{\Sigma^0 \Lambda}}
\def\es{\!\!\! &=& \!\!\!}
\def\ap{\!\!\! &\approx& \!\!\!}
\def\ar{&+& \!\!\!}
\def\ek{&-& \!\!\!}
\def\kek{\!\!\!&-& \!\!\!}
\def\cp{&\times& \!\!\!}
\def\se{\!\!\! &\simeq& \!\!\!}
\def\eqv{&\equiv& \!\!\!}
\def\kpm{&\pm& \!\!\!}
\def\kmp{&\mp& \!\!\!}
\def\mcdot{\!\cdot\!}
\def\erar{&\rightarrow&}


\def\simlt{\stackrel{<}{{}_\sim}}
\def\simgt{\stackrel{>}{{}_\sim}}


\renewcommand{\textfraction}{0.2}    
\renewcommand{\topfraction}{0.8}   

\renewcommand{\bottomfraction}{0.4}   
\renewcommand{\floatpagefraction}{0.8}
\newcommand\mysection{\setcounter{equation}{0}\section}

\def\baeq{\begin{appeq}}     \def\eaeq{\end{appeq}}  
\def\baeeq{\begin{appeeq}}   \def\eaeeq{\end{appeeq}}
\newenvironment{appeq}{\beq}{\eeq}   
\newenvironment{appeeq}{\beeq}{\eeeq}
\def\bAPP#1#2{
 \markright{APPENDIX #1}
 \addcontentsline{toc}{section}{Appendix #1: #2}
 \medskip
 \medskip
 \begin{center}      {\bf\LARGE Appendix #1 :}{\quad\Large\bf #2}
\end{center}
 \renewcommand{\thesection}{#1.\arabic{section}}
\setcounter{equation}{0}
        \renewcommand{\thehran}{#1.\arabic{hran}}
\renewenvironment{appeq}
  {  \renewcommand{\theequation}{#1.\arabic{equation}}
     \beq
  }{\eeq}
\renewenvironment{appeeq}
  {  \renewcommand{\theequation}{#1.\arabic{equation}}
     \beeq
  }{\eeeq}
\nopagebreak \noindent}

\def\eAPP{\renewcommand{\thehran}{\thesection.\arabic{hran}}}

\renewcommand{\theequation}{\arabic{equation}}
\newcounter{hran}
\renewcommand{\thehran}{\thesection.\arabic{hran}}

\def\bmini{\setcounter{hran}{\value{equation}}
\refstepcounter{hran}\setcounter{equation}{0}
\renewcommand{\theequation}{\thehran\alph{equation}}\begin{eqnarray}}
\def\bminiG#1{\setcounter{hran}{\value{equation}}
\refstepcounter{hran}\setcounter{equation}{-1}
\renewcommand{\theequation}{\thehran\alph{equation}}
\refstepcounter{equation}\label{#1}\begin{eqnarray}}


\newskip\humongous \humongous=0pt plus 1000pt minus 1000pt
\def\caja{\mathsurround=0pt}


\title{
         {\Large
                 {\bf
Spin--3/2 to spin--1/2 heavy baryons and pseudoscalar mesons transitions  in
QCD              }
         }
      }

\author{\vspace{1cm}\\
{\small T. M. Aliev \thanks {e-mail:
taliev@metu.edu.tr}~\footnote{permanent address: Institute of
Physics,Baku,Azerbaijan}\,\,, K. Azizi \thanks {e-mail:
kazizi@dogus.edu.tr}\,\,, M. Savc{\i} \thanks
{e-mail: savci@metu.edu.tr}} \\
{\small Physics Department, Middle East Technical University,
06531 Ankara, Turkey }\\
{\small$^\ddag$ Physics Division,  Faculty of Arts and Sciences,
Do\u gu\c s University,} \\
{\small Ac{\i}badem-Kad{\i}k\"oy,  34722 Istanbul, Turkey}}

\date{}

\begin{titlepage}
\maketitle
\thispagestyle{empty}

\begin{abstract}
The strong coupling constants of light pseudoscalar mesons with spin--3/2 and
spin--1/2 heavy baryons are calculated in the  framework of light cone QCD sum rules.
It is shown that each class of transitions among  members of the sextet spin--3/2 to
sextet spin--1/2 baryons and that of the sextet spin--3/2 to 
spin--1/2 anti--triplet baryons is 
 described by only one invariant function. We also
estimate the widths of kinematically allowed transitions. Our results on decay
widths are in good agreement with the existing experimental data, as well as
 predictions of other nonperturbative approaches. 
\end{abstract}

~~~PACS number(s): 11.55.Hx, 13.75.Gx, 13.75.Jz
\end{titlepage}

\section{Introduction}

Theoretical and experimental studies of the flavored 
hadrons are among the most promising areas in particle physics. 
From theoretical point of view, this can
be explained by the fact that the heavy flavored baryons provide a rich
laboratory to study  predictions of the heavy quark effective theory.
On the other hand, these baryons have many weak and strong decay channels and
therefore measurement of these channels can give essential information about
the quark structure of heavy baryons. During the last decade, highly exciting
experimental results have been obtained on the spectroscopy of heavy hadrons. All
ground states of heavy hadrons with $c$ quark have been observed  \cite{Rh32v01}.
The new states of heavy baryons are also discovered in BaBar, BELLE, CDF
and D$\rlap/$0 Collaborations. The operation of LHC will open a new window
for more detailed investigation of these new baryons \cite{Rh32v02}.

At present, we have experimental information on the strong one--pion decays
for the $\Sigma_c$ \cite{Rh32v03,Rh32v04,Rh32v05}, $\Sigma_c^*$
\cite{Rh32v04,Rh32v06} and $\Xi_c^*$ \cite{Rh32v07,Rh32v08} baryons.
The strong coupling constants of pseudoscalar mesons with heavy baryons are
the main unknown parameters of these transitions. Therefore, a reliable
estimation of these strong coupling constants in the framework of QCD receives great
interest. At hadronic scale, the strong coupling constant $\alpha_s(Q^2)$ is
large and hence perturbative theory becomes invalid. For this reason,
estimation of the coupling constants becomes impossible starting from the 
fundamental QCD Lagrangian and some nonperturbative methods are needed. Among
many nonperturbative approaches, the QCD sum rule \cite{Rh32v09} is one of
the most powerful method in studying the properties of hadrons. The main
advantage of this method is that, it is based on fundamental QCD Lagrangian.
In the present work, we estimate the strong coupling constants of
pseudoscalar mesons in the transitions of spin--3/2 to spin--1/2 heavy
baryons within light cone QCD sum rules method (for more about this method,
see \cite{Rh32v10}). Note that the strong coupling
constants of pseudoscalar and vector mesons with heavy baryons in the
spin--1/2 to spin--1/2 transitions are studied in \cite{Rh32v11} and
\cite{Rh32v12}. 

The rest of the paper is organized as follows.
In section 2, the light cone sum
rules for the coupling constants of pseudoscalar mesons with heavy baryons
in spin--3/2 to spin--1/2 transitions are calculated. In section 3, the
numerical analysis of the obtained sum rules is performed and a comparison
of our results with the predictions of other approaches as well as 
existing experimental results is made. 

\section{Light cone QCD sum rules for the pseudoscalar mesons with heavy
baryons in spin--3/2 to spin--1/2 transitions}

In this section, the strong coupling constants of light pseudoscalar mesons
with heavy baryons in spin--3/2 to spin--1/2 transitions are calculated.
Before making an attempt in estimating these coupling constants, few words
about $SU(3)_f$ classification of heavy baryons are in order. Heavy baryons
with a single heavy quark and two light quarks can be decomposed into two
multiplets, namely, sextet $6_F$ and anti--triplet $\bar{3}_F$ due to the
symmetry property of flavor and color structures of these baryons. This
observation leads to the result of total spin $J^P = (3/2)^+$ or $(1/2)^+$ 
for $6_F$ and $J^P = (1/2)^+$ for $\bar{3}_F$.
In the present work, we consider $J=3/2$ states in $6_F$ and investigate 
sextet to sextet (S$^*$SP) and sextet to 
anti--triplet (S$^*$AP) transitions  with the participation of light
pseudoscalar mesons, where $S^*$, $S$ and $A$ stand for sextet with spin--3/2, 
sextet with spin--1/2  and anti--triplet spin--1/2 states, respectively.

We now pay our attention to the calculation of the strong coupling constants
of pseudoscalar mesons with heavy baryons in spin--3/2 to spin--1/2
transitions. To derive the light cone sum rules for $S^* SP$ and $S^* AP$
transitions we consider the following general correlation function:
\bea
\label{eh32v01}
\Pi_\mu^{(i)} = i \int d^4x e^{ipx} \lla P(q) \vel 
\eta^{(i)} (x) \bar{\eta}_\mu (0) \ver 0 \rra~,
\eea
where $\eta^{(i)}(x)$ are the interpolating currents of the heavy baryons
with spin--1/2 in sextet $(i=1)$ and anti--triplet  $(i=2)$ representation and
$\bar{\eta}_\mu$ is the interpolating current for the sextet $J^P=3^+/2$
states.
The correlation function (\ref{eh32v01}) can be calculated in terms of
hadrons (phenomenological part) and in terms of quark--gluon degrees of
freedom in deep Euclidean region, i.e., when $p^2 \to -\infty$. Equating then
these representations of the correlation function using the dispersion
relation, we get the sum rules for strong coupling constants of light
pseudoscalar mesons with heavy baryons.

We proceed by calculating the phenomenological part of the correlation function.
The expression for the phenomenological part is obtained by saturating it
with the full set of hadrons carrying the same quantum numbers as the
corresponding interpolating current.  Isolating the contributions of the 
ground state baryons, one can easily obtain
\bea
\label{eh32v02}
\Pi_\mu =  {\lla 0 \vel \eta^{(i)} (x) \ver
B_(p) \rra \lla B(p) P(q) \vel \right.
B^*(p+q) \rra \lla B^*(p+q) \vel \bar{\eta}_\mu (0)\ver
0 \rra \over \ga p^2-m_2^2 \dr \left[(p+q)^2-m_1^2\right]}
+ \cdots~,
\eea
where $m_1$ and $m_2$ are the masses of the initial and final heavy baryons,
$p+q$ and $p$ represent their four--momentum, respectively, and dots
represent contributions coming from higher states and continuum.
It follows from Eq. (\ref{eh32v02}) that in obtaining the phenomenological
part of the correlation function, the matrix elements, $\lla 0 \vel
\eta^{(i)} (x) \ver B_(p)\rra$, $\lla B(p) P(q) \vel \right.
B(p+q) \rra$ and $\lla B(p+q) \vel \bar{\eta}_\mu (0)\ver
0 \rra$ are needed. These matrix elements are determined as follows:
\bea
\label{eh32v03}
\lla 0 \vel \eta \ver B(p) \rra \es \lambda_2^{(i)} u(p)~, \nnb \\
\lla B^*(p+q) \vel \eta_\mu \ver 0 \rra \es \lambda_1 \bar{u}_\mu(p+q)~, \nnb \\
\lla P(q) B(p) \vel \right. B^*(p+q) \rra \es g_{B^*BP} \bar{u} (p) u_\alpha (p+q)
q^\alpha~,
\eea
where $\lambda_2$ and $\lambda_1$ are the residues of spin--1/2 and
spin--3/2 heavy baryons, respectively, $g$ is the coupling constant of heavy
baryons with pseudoscalar mesons, and $u_\mu$ is the Rarita--Schwinger
spinor. Using Eqs. (\ref{eh32v03}) and (\ref{eh32v02}) and performing
summation over spins of spin--1/2 and spin--3/2 baryons,
\bea
\label{eh32v04}
\sum_s u(p,s) \bar{u}(p,s) \es (\rlap/{p} + m_2)~,\nnb \\
\sum_s u_\alpha (p+q,s) \bar{u}_\mu(p+q,s) \es - (\rlap/{p} + \rlap/{q} + m_1)
\Bigg\{ g_{\alpha\mu} - {\gamma_\alpha \gamma_\mu \over 3} -
{2 (p+q)_\alpha (p+q)_\mu \over 3 m_1^2} \nnb \\ 
\ar {(p+q)_\alpha \gamma_\mu - (p+q)_\mu \gamma_\alpha \over 3 m_1}\Bigg\}~,
\eea
in principle, one can obtain the expression for the phenomenological part of
the correlation function. But at this point appear two unpleasant
problems: a)  The spin--1/2 baryons also 
contribute to the matrix element $\lla 0 \vel
\eta_\mu \ver B(p,3/2) \rra$ of spin--3/2 baryons (see also \cite{Rh32v11}). Indeed,
\bea
\label{eh32v05}
\lla 0 \vel \eta_\mu \ver B(p,1/2) \rra = A \Bigg( \gamma_\mu - {4\over m_2}
p_\mu \Bigg) u(p)~,
\eea
hence, the current $\eta_\mu$ couples to both  spin--3/2 and spin--1/2 states.
Using Eqs. (\ref{eh32v02}), (\ref{eh32v04}) and (\ref{eh32v05}), one can see
that the unwanted contributions coming from spin--1/2 states contain
structures proportional to $\gamma_\mu$ at the far right end or 
$(p+q)_\mu$. b) The second problem is related to the fact that the
structures which appear in the phenomenological part of the correlation
function are not all independent. Both these problems can be removed by
ordering the Dirac matrices in a specific form. In this work, the
Dirac matrices are ordered in the form $\rlap/{q}\rlap/{p}{\gamma}_\mu$
and the coefficient of the structure $\rlap/{q}q_\mu$ is chosen 
in order to calculate 
the aforementioned strong coupling constant, which is free of the spin--1/2
contributions. Using the ordering procedure, we get the following representation for the coefficient of the selected structure in the phenomenological part:
\bea
\label{eh32v06}
\Pi_\mu^{(i)} = {g \lambda_1 \lambda_2^{(i)} m_2 \over
[m_1^2-(p+q)^2] (m_2^2 - p^2)} + \mbox{\rm other structures}.
\eea

In order to obtain  sum rules for the coupling constant appearing in Eq. 
(\ref{eh32v06}), we need to calculate the correlation function also from the QCD
side. Before calculating it, we shall first find the relations among the
correlation functions corresponding to different transition channels. In
more concrete words, we shall find the relations among the coefficient functions of
the structure $\rlap/{q}q_\mu$ for different transition channels. For this
purpose we follow an approach whose main ingredients are presented in
\cite{Rh32v14,Rh32v15,Rh32v16,Rh32v17}. 

In obtaining the relations among the correlation functions describing
various spin--3/2 to spin--1/2 heavy baryon transitions, as well as, in
obtaining the  theoretical part of QCD sum rules, the forms of the interpolating
currents are needed. In constructing the interpolating currents, we
will use the fact that the interpolating currents for the particles in
sextet (anti--triplet) representations should be symmetric (antisymmetric)
with respect to the light quarks. Using this fact, the interpolating current for
baryons in sextet representation with $J=3/2$ can be written as:
\bea
\label{eh32v07}
\eta_\mu = A \epsilon^{abc} \Big\{ (q_1^a C \gamma_\mu q_2^b) Q^c + (q_2^a C
\gamma_\mu Q^b) q_1^c + (Q^a C \gamma_\mu q_1^b) q_2^c \Big\}~,
\eea
where $A$ is the normalization factor, $a$, $b$ and $c$ are the color indices. In
Table 1, we present the values of $A$ and light quark content of heavy spin-3/2
baryons.
\begin{table}[thb]

\renewcommand{\arraystretch}{1.3}
\addtolength{\arraycolsep}{-0.5pt}
\small
$$
\begin{array}{|l|c|c|c|c|c|c|}
\hline \hline
 & \Sigma_{b(c)}^{*+(++)} & \Sigma_{b(c)}^{*0(+)} & \Sigma_{b(c)}^{*-(0)}  
 & \Xi_{b(c)}^{*0(+)}    & \Xi_{b(c)}^{*-(0)} 
 & \Omega_{b(c)}^{*-(0)}          \\  \hline
 q_1 & u & u & d & u & d & s \\
 q_2 & u & d & d & s & s & s  \\
 A   & \sqrt{1/3} & \sqrt{2/3} & \sqrt{1/3}
     & \sqrt{2/3} & \sqrt{2/3} & \sqrt{1/3} \\
\hline \hline
\end{array}
$$
\caption{The light quark content $q_1$ and $q_2$ for the sextet baryons with
spin--3/2}
\renewcommand{\arraystretch}{1}
\addtolength{\arraycolsep}{-1.0pt}
\end{table}
The general
form of the interpolating currents for the heavy spin-1/2 sextet
and antitriplet baryons can be written as ( for example see  \cite{Rh32v18}):
\bea
\label{estp05}
\eta_Q^{(s)} \es - {1\over \sqrt{2}} \epsilon^{abc} \Big\{ \Big( q_1^{aT} 
C Q^b \Big) \gamma_5 q_2^c + \beta \Big( q_1^{aT} C \gamma_5 Q^b \Big) q_2^c -
\Big[\Big( Q^{aT} C q_2^b \Big) \gamma_5 q_1^c + \beta \Big( Q^{aT} C
\gamma_5 q_2^b \Big) q_1^c \Big] \Big\}~, \nnb\\
\label{estp06} 
\eta_Q^{(a)} \es {1\over \sqrt{6}} \epsilon^{abc} \Big\{ 2 \Big( q_1^{aT} 
C q_2^b \Big) \gamma_5 Q^c + 2 \beta \Big( q_1^{aT} C \gamma_5 q_2^b \Big) Q^c
+ \Big( q_1^{aT} C Q^b \Big) \gamma_5 q_2^c + \beta \Big(q_1^{aT} C
\gamma_5 Q^b \Big) q_2^c \nnb \\
\ar \Big(Q^{aT} C q_2^b \Big) \gamma_5 q_1^c +
\beta \Big(Q^{aT} C \gamma_5 q_2^b \Big) q_1^c \Big\}~,
\eea
where $\beta$ is an arbitrary constant and $\beta=-1$ corresponds to the
Ioffe current and superscripts $s$ and $a$ stand for symmetric and antisymmetric spin--1/2 currents, respectively. The light quark content of the heavy baryons with spin--1/2 in the sextet and 
anti--triplet representations are given in Table 2.


\begin{table}[thb]

\renewcommand{\arraystretch}{1.3}
\addtolength{\arraycolsep}{-0.5pt}
\small
$$
\begin{array}{|l|c|c|c|c|c|c|c|c|c|}
\hline \hline
  & \Sigma_{b(c)}^{+(++)} & \Sigma_{b(c)}^{0(+)}  & \Sigma_{b(c)}^{-(0)} 
  & \Xi_{b(c)}^{-(0)'}    & \Xi_{b(c)}^{0(+)'}
  & \Omega_{b(c)}^{-(0)}  & \Lambda_{b(c)}^{0(+)}
  & \Xi_{b(c)}^{-(0)}     & \Xi_{b(c)}^{0(+)}    \\  \hline
q_1 & u & u & d & d & u & s & u & d & u \\
q_2 & u & d & d & s & s & s & d & s & s \\
\hline \hline
\end{array}
$$
\caption{The light quark content $q_1$ and $q_2$ for the sextet and
anti--triplet baryons with spin--1/2}
\renewcommand{\arraystretch}{1}
\addtolength{\arraycolsep}{-1.0pt}
\end{table}
  

After introducing the explicit expressions for the interpolating currents, we
are ready now to obtain the relations among the correlation functions that
describe different transitions. It should be noted here that the relations
which are presented below are independent of the choice of  structures,
while the expressions of the correlation functions are all structure
dependent. 

In order to obtain the relations among the correlation functions responsible
for different transitions, we consider the $\Sigma_{b}^{*0} \to
\Sigma_{b}^{0} \pi^0$ and $\Sigma^{*0} \to \Lambda \pi^0$ 
transitions which describe sextet spin--3/2 to sextet spin--1/2
and sextet spin--3/2  to anti--triplet spin--1/2 transitions, respectively. These invariant
functions can be written in the general form as
\bea
\label{eh32v09}
\Pi^{\Sigma_{b}^{*0} \to \Sigma_{b}^{0} \pi^0} \es 
g_{\pi^0 \bar{u}u} \Pi_1^{(1)}(u,d,b) +
g_{\pi^0 \bar{d}d} \Pi_1^{(1)'}(u,d,b) +
g_{\pi^0 \bar{b}b} \Pi_2^{(1)'}(u,d,b)~,\nnb \\
\Pi^{\Sigma^{*0} \to \Lambda \pi^0} \es 
g_{\pi^0 \bar{u}u} \Pi_1^{(2)}(u,d,b) +
g_{\pi^0 \bar{d}d} \Pi_1^{(2)'}(u,d,b) +
g_{\pi^0 \bar{b}b} \Pi_2^{(2)'}(u,d,b)~,
\eea
where superscripts (1) and (2) correspond to S$^*$SP and 
S$^*$AP transitions, respectively. The interpolating current for
$\pi^0$ is written as:
\bea
\label{nolabel}
J_{\pi^0} = \sum g_{\pi^0\bar{q}q} \bar{q}\gamma_5q  = {1\over \sqrt{2}} (\bar{u}
\gamma_5 u - \bar{d} \gamma_5 d)~.
\eea
It follows from this expression that,
\bea
\label{eh32v10}
g_{\pi^0\bar{u}u} = - g_{\pi^0\bar{d}d} = {1\over \sqrt{2}}~,
~~g_{\pi^0\bar{b}b} = 0~.
\eea
The invariant functions $\Pi_1^{(i)}(u,d,b)$, $\Pi_1^{(i)'}(u,d,b)$, and
$\Pi_2^{(i)}(u,d,b)$ describe the radiation of $\pi^0$ from $u$, $d$ and $b$
quarks, respectively, and they are formally defined in the following way:
\bea         
\label{nolabel}
\Pi_1^{(i)}(u,d,b) \es \lla \bar{u} u \vel \Sigma_{b}^{*0} \Sigma_{b}^{0}
(\Lambda_b) \ver 0 \rra~, \nnb \\
\Pi_1^{(i)'}(u,d,b) \es \lla \bar{d} d \vel \Sigma_{b}^{*0} \Sigma_{b}^{0}   
(\Lambda_b) \ver 0 \rra~, \nnb \\
\Pi_2^{(i)}(u,d,b) \es \lla \bar{b} b \vel \Sigma_{b}^{*0} \Sigma_{b}^{0}   
(\Lambda_b) \ver 0 \rra~. 
\eea

Remembering the fact that the interpolating currents for sextet spin--3/2 and
sextet spin--1/2 baryons are symmetric with respect to the exchange of light
quarks, while the interpolating currents for spin--1/2 anti--triplet baryons
are antisymmetric, we can write,
\bea
\Pi_1^{(1)'}(u,d,b) \es \Pi_1^{(1)}(d,u,b)~, \nnb \\
\Pi_1^{(2)'}(u,d,b) \es -\Pi_1^{(2)}(d,u,b)~. 
\eea
Using these relations and Eqs. (\ref{eh32v09}) and (\ref{eh32v10}), we get,
\bea
\label{eh32v11}
\Pi^{\Sigma_{b}^{*0} \to \Sigma_{b}^{0} \pi^0 (\Sigma^{*0} \to \Lambda
\pi^0)} = {1\over \sqrt{2}} \Big[ \Pi_1^{(i)} (u,d,b) \mp \Pi_1^{(i)}
(d,u,b) \Big]~,
\eea
where $i=1\,(i=2)$ and upper (lower) sign describes S$^*$SP (S$^*$AP) transition.

The invariant function responsible for the $\Xi_b^{*0} \to \Xi_b^{0'} \pi^0$
and $\Xi_b^{*0} \to \Xi_b^0 \pi^0$
transitions can be obtained from the $\Sigma_{b}^{*0} \to \Sigma_{b}^{0}
\pi^0$ and $\Sigma^{*0}_b \to \Lambda_b^0 \pi^0$
channels by noting that the interpolating currents for $\Xi_b^{*0}$, $\Xi_b^{0'}$ 
and $\Xi_b^0$ can be obtained from the one for  $\Sigma^{*0}_b$, $\Sigma^{0}_b$ 
and $\Lambda_b^0$ by making the replacement $d \to s$, and taking into account
the fact that $g_{\pi^0 \bar{s}s}=0$. As a result, we get,
\bea           
\label{eh32v12}
\Pi^{\Xi_b^{*0} \to \Xi_b^{0'} \pi^0 (\Xi_b^{*0} \to \Xi_b^{0} \pi^0)} = 
{1\over \sqrt{2}} \Pi_1^{(i)}(u,s,b)~.
\eea

The invariant functions corresponding to $\Xi_b^{*-} \to \Xi_b^{-'} \pi^0$
and $\Xi_b^{*-} \to \Xi_b^- \pi^0$
transitions can be obtained from the  $\Xi_b^{*0} \to \Xi_b^{0'} \pi^0$
and $\Xi_b^{*0} \to \Xi_b^0 \pi^0$ channels
with the help of the replacement $u \to d$, as a result of which we get,
\bea           
\label{eh32v13}
\Pi^{\Xi_b^{*-} \to \Xi_b^{-'} \pi^0(\Xi_b^{*-} \to \Xi_b^- \pi^0)} = 
- {1\over \sqrt{2}} \Pi_1^{(i)}(d,s,b)~.
\eea

Calculation of the coupling constants of the sextet spin--3/2 to
sextet and anti--triplet spin--1/2 transitions with other pseudoscalar 
mesons can be done in
a similar way as for the $\pi^0$ meson. Note that in our calculations, the 
mixing between $\eta$ and
$\eta^{'}$ mesons is neglected and the interpolating current of $\eta$
meson has the following form:
\bea             
\label{nolabel}
J_\eta = {1\over \sqrt{6}} [ \bar{u}\gamma_5 u + \bar{d}\gamma_5 d
- 2 \bar{s}\gamma_5 s]~,
\eea 
that gives,
\bea             
\label{nolabel}
g_{\eta\bar{u}u} \es g_{\eta\bar{d}d} = {1\over\sqrt{6}}~,~\mbox{\rm and} 
~~g_{\eta\bar{s}s} = -{2\over \sqrt{6}}~. 
\eea
Using this expression let us  consider, for example, 
the $\Sigma^{*0}_b \to \Sigma_b^0 \eta$ and $\Sigma^{*0}_b \to \Lambda_b^0 \eta$
transitions. Following the same lines of calculations as in the $\pi^0$ meson
case, we immediately get,
\bea             
\label{eh32v14} 
\Pi^{\Sigma^{*0}_b \to \Sigma_b^0 \eta(\Sigma^{*0}_b \to \Lambda_b^0 \eta)} = 
{1\over \sqrt{6}} [\Pi_1^{(i)}(u,d,b) \pm \Pi_1^{(i)}(d,u,b)]~.
\eea             

The invariant function responsible for the $\Xi_b^{*0} \to \Xi_b^{0'} \eta$ and 
$\Xi_b^{*0} \to \Xi_b^0 \eta$
transition can be written as:
\bea
\label{eh32v15}
\Pi^{\Xi_b^{*0} \to \Xi_b^{0'} \eta (\Xi_b^{*0} \to \Xi_b^0 \eta)} \es 
g_{\eta\bar{u}u} \Pi_1^{(i)}(u,s,b) +
 g_{\eta\bar{s}s} \Pi_1^{(i)'}(u,s,b) +  g_{\eta\bar{b}b} \Pi_2^{(i)}(u,s,b) \nnb \\
\es {1\over \sqrt{6}} [\Pi_1^{(i)}(u,s,b) - 2 \Pi_1^{(i)'}(u,s,b)] \nnb \\
\es {1\over \sqrt{6}} [\Pi_1^{(i)}(u,s,b) \mp 2 \Pi_1^{(i)}(s,u,b)]~.
\eea
The relations among invariant functions involving charged pseudoscalar
$\pi^\pm$ mesons can be obtained from previous results by taking  into account the 
following arguments. For instance, let us consider the
 $\Sigma^{*+}_b \to \Lambda_b^0 \pi^+$ transition.
In the $\Sigma^{*0}_b \to \Lambda_b^0
\pi^0$ transition, the $u(d)$ quark from $\Sigma^{*0}_b$ and $\Lambda_b^0$
baryons forms the final $\bar{u}u(\bar{d}d)$ state, and the $d(u)$ and $b$ quarks
behave like spectators. In the case of charged $\pi^+$ meson, the $d$ quark from
$\Lambda_b^0$ and $u$ quark from $\Sigma^{*0}_b$ form the $\bar{u}d$ final
state, and the remaining $d(u)$ and $b$ quarks are the spectators. For this
reason, one can expect that these two matrix elements should be proportional
to each other and explicit calculations show that this indeed is the case.
Hence,
\bea           
\label{eh32v16}
\Pi^{\Sigma^{*+}_b \to \Lambda_b^0 \pi^+} \es \lla \bar{u}d \vel \Sigma^{*+}_b
\Lambda_b^0 \ver 0 \rra = \sqrt{2} \lla \bar{d}d \vel \Sigma^{*0}_b 
\Lambda_b^0 \ver 0 \rra \nnb \\
\es - \sqrt{2} \Pi_1^{(2)}(d,u,b)~.
\eea
making the replacement $u \lrar d$ in Eq. (\ref{eh32v16}), we get
\bea           
\label{eh32v17}
\Pi^{\Sigma^{*-}_b \to \Lambda_b^0 \pi^-} \es \sqrt{2} \Pi_1^{(2)}(u,d,b)~.
\eea
All remaining relations among the invariant functions responsible for the
spin--3/2 and spin--1/2 transitions involving pseudoscalar mesons are
presented in the Appendix.

After establishing the relations among the invariant functions, we now
proceed by calculating the invariant functions from QCD side in deep
Euclidean region $-p^2 \to \infty$, $-(p+q)^2 \to \infty$, using the
operator product expansion (OPE). The main nonperturbative input parameters
in the calculation of the theoretical part of the correlation function  are the distribution amplitudes (DAs) of the pseudoscalar mesons.
These (DAs) of the pseudoscalar mesons are involved in determining the
matrix elements of the nonlocal operators between the vacuum and one
pseudoscalar meson states, i.e., $\lla p(q) \vel \bar{q} (x) \Gamma q(0)
\ver 0 \rra$ and $\lla p(q) \vel \bar{q} (x) G_{\mu\nu} q(0) \ver 0 \rra$,
where $\Gamma$ is any Dirac matrix. The DAs of pseudoscalar mesons  up to twist--4 accuracy are 
given in \cite{Rh32v19}.

In the calculation of the theoretical part of the correlation function, we
also need to know the expressions of the light and heavy quark propagators.
The light quark propagator, in presence of an external field, is calculated
in \cite{Rh32v21}:
\bea
\label{eh32v18}
S_q(x) \es {i \rlap/x\over 2\pi^2 x^4} - {m_q\over 4 \pi^2 x^2} -
{\lla \bar q q \rra\over 12} \left(1 - i {m_q\over 4} \rlap/x \right) -
{x^2\over 192} m_0^2 \lla \bar q q \rra  \left( 1 -
i {m_q\over 6}\rlap/x \right) \nnb \\
&&  - i g_s \int_0^1 du \left[{\rlap/x\over 16 \pi^2 x^2} G_{\mu \nu} (ux)
\sigma_{\mu \nu} - {i\over 4 \pi^2 x^2} u x^\mu G_{\mu \nu} (ux) \gamma^\nu
\right. \nnb \\
&& \left.
 - i {m_q\over 32 \pi^2} G_{\mu \nu} \sigma^{\mu
 \nu} \left( \ln \left( {-x^2 \Lambda^2\over 4} \right) +
 2 \gamma_E \right) \right]~,
\eea
where $\gamma_E \simeq 0.577$ is the Euler constant, and $\Lambda$ is the
scale parameter. In further numerical calculations, we choose it as
$\Lambda=(0.5 \div 1)~GeV$ (see \cite{Rh32v22,Rh32v23}).
The heavy quark propagator in an external field has the following form:
\bea
\label{eh32v19}
S_Q(x) = S_Q^{free}(x) -
ig_s \int {d^4k \over (2\pi)^4} e^{-ikx} \int_0^1
du \Bigg[ {\rlap/k+m_Q \over 2 (m_Q^2-k^2)^2} G^{\mu\nu} (ux)
\sigma_{\mu\nu} +
{u \over m_Q^2-k^2} x_\mu G^{\mu\nu} \gamma_\nu \Bigg]~,
\eea
where $S_Q^{free}(x)$ is the free heavy quark operator in x--representation, 
which is given by:
\bea
\label{nolabel}
S_Q^{free} (x) = 
{m_Q^2 \over 4 \pi^2} {K_1(m_Q\sqrt{-x^2}) \over \sqrt{-x^2}} -
i {m_Q^2 \rlap/{x} \over 4 \pi^2 x^2} K_2(m_Q\sqrt{-x^2})~, 
\eea
where $K_1$ and $K_2$ are the modified Bessel function of the second kind.

Using the explicit expressions of the heavy and light quark propagators, as
well as, definition of the DAs of the pseudoscalar mesons, the correlation
function can be calculated from the QCD side. Choosing the coefficient of
the structure $\rlap/{q}q_\mu$ from both sides of the correlation function
and applying double Borel
transformations with respect to the variables $-p^2$ and $-(p+q)^2$, in order
to suppress the contributions of higher states and continuum, we get the sum
rules for the strong coupling constants of pseudoscalar mesons with sextet
spin--3/2 and spin--1/2 heavy baryons as:  
\bea
\label{eh32v20}
g = {1 \over \lambda_1 \lambda_2^{(i)} m_2}  
e^{{m_1^2 \over M_1^2} + {m_2^2 \over M_2^2} + 
{m_P^2 \over M_1^2 + M_2^2}}\, \Pi_1^{(i)}~,
\eea
where $M_1^2$ and $M_2^2$ are the Borel masses in the
initial and final channels, respectively. The masses of initial and final
heavy baryons are very close to each other, so that we can choose
$M_1^2=M_2^2 = 2 M^2$. The residues $\lambda_1$ of spin--3/2 and $\lambda_2$
of spin--1/2 are calculated in \cite{Rh32v24}. The explicit expressions for $\Pi_1^{(i)}$
are quite lengthy, so as an example, we present only the  $\Pi_1^{(1)}$, which is obtained as:

\bea
%
%
&&e^{m_Q^2/M^2 - m_{\cal P}^2/M^2} \Pi_1^{(1)} (u,d,b) = \nnb \\
\ek {(1+\beta)\over 8 \sqrt{6} \pi^2} M^4 \mu_{\cal P} \Big\{
2 i_2({\cal T},1) - m_Q^2 \Big[ i_2({\cal T},1) - i_2({\cal T},v)\Big] I_2 \Big\} \nnb \\
\ar {(1-\beta)\over 16 \sqrt{6} \pi^2} M^4 m_Q^3
\Big\{ 2 f_{\cal P} \phi_\eta(u_0) I_2 +
m_Q \Big( \mu_{\cal P} \Big[ i_3({\cal T},1) - 2 i_3({\cal T},v)\Big] - 2 f_{\cal P}
m_Q \phi_\eta(u_0) \Big) I_3 \Big\} \nnb \\
\ar {1\over 96 \sqrt{6} \pi^2} M^4 \mu_{\cal P} \Big\{ 12 (1+3 \beta)
i_2({\cal T},v) +
12 \beta m_Q^2 \Big[i_3({\cal T},1) - i_3({\cal T},v)\Big] I_2 \nnb \\
\ek \beta m_Q^4 \Big[ 
\Big(6 \phi_{\cal P}(u_0) + (1-\widetilde{\mu}_{\cal P}^2) [4 \phi_\sigma(u_0) -
\phi_\sigma^\prime(u_0)] \Big) I_3 - m_Q^2 \Big( 6 \phi_{\cal P}(u_0) - 
(1-\widetilde{\mu}_{\cal P}^2)    
\phi_\sigma^\prime(u_0)\Big) I_4
\Big]\Big\} \nnb \\
%
%
\ek {1 \over 16 \sqrt{6} \pi^2} M^2 m_Q m_{\cal P}^2 f_{\cal P}   
\Big\{4 i_1({\cal A}_\parallel,1) + 4  i_1({\cal A}_\perp,1) -  
 i_2({\cal A}_\parallel,1) - 2 i_2({\cal V}_\perp,1) \nnb \\
\ek \beta \Big[4 i_1({\cal V}_\parallel,1) + 4 i_1({\cal V}_\perp,1) +  
i_2({\cal A}_\parallel,1) - 4 i_2({\cal A}_\perp,1) - 
2 i_2({\cal V}_\parallel,1) + 2 i_2({\cal V}_\perp,1) \Big] \Big\} I_1 \nnb \\
\ar {1 \over 32 \sqrt{6} \pi^2} M^2 m_Q^2 m_{\cal P}^2 
\Big\{-4 \beta m_Q f_{\cal P} i_2({\cal A}_\parallel,1) +
8 \beta m_Q f_{\cal P} i_2({\cal A}_\perp,1) +
9 \beta \mu_{\cal P} i_2({\cal T},1) \nnb \\
\ek 2 (1-3 \beta) m_Q f_{\cal P}
i_2({\cal V}_\parallel,1) - 4 (1+\beta) m_Q f_{\cal P} i_2({\cal V}_\perp,1)
-  4 (1-\beta) m_Q f_{\cal P} i_2({\cal A}_\parallel,v) \nnb \\
\ar \mu_{\cal P}(1 - 5 \beta) i_2({\cal T},v) - m_Q f_{\cal P}
\Big[(1-\beta) \mathbb{A}(u_0) + 2 (1+\beta) \widetilde{i}_4({\Bbb{B}})\Big]\Big\}
I_2 \nnb \\
\ar {1 \over 12 \sqrt{6} \pi^2} M^2 f_{\cal P} \Big\{ 
- 3(1+\beta) m_Q m_{\cal P}^2 \Big[ i_1({\cal A}_\parallel,1) +
i_1({\cal A}_\perp,1) + i_1({\cal V}_\parallel,1) + 
i_1({\cal V}_\perp,1) \nnb \\
\ek 2 \Big( i_1({\cal A}_\parallel,v) +
i_1({\cal A}_\perp,v) \Big)\Big] + 2 (2 + \beta) \pi^2 \dd
\phi_\eta(u_0)\Big\} \nnb \\
\ar {1 \over 48 \sqrt{6} \pi^2} M^2 m_Q^6 m_{\cal P}^2 \mu_{\cal P}
(1-\widetilde{\mu}_{\cal P}^2) \beta \phi_\sigma(u_0) I_4 \nnb \\
\ar {1 \over 96 \sqrt{6} \pi^2} M^2 m_Q^4 m_{\cal P}^2
\Big\{15 (1-\beta) \mu_{\cal P} i_2({\cal T},1) - 
6 (5-2\beta) \mu_{\cal P} i_2({\cal T},v) \nnb \\
\ar 6 (1+\beta) m_Q f_{\cal P} \widetilde{i}_4({\Bbb{B}}) -
2 \beta \mu_{\cal P} (1-\widetilde{\mu}_{\cal P}^2) \phi_\sigma(u_0)\Big\}
I_3 \nnb \\
%
%
\ar{1\over 96 \sqrt{6} M^6} m_Q^4 m_{\cal P}^2 m_0^2 f_{\cal P} \dd 
\Big\{(2+\beta) \mathbb{A}(u_0) + 8 \Big[ i_1({\cal A}_\parallel,1) +      
i_1({\cal A}_\perp,1) \nnb \\
\ar \beta i_1({\cal V}_\parallel,1) +            
\beta i_1({\cal V}_\perp,1) - 2 i_1({\cal A}_\parallel,v) -          
2 i_1({\cal A}_\perp,v)\Big] \Big\} \nnb \\
%
%
\ek {1\over 192 \sqrt{6} M^4} m_Q^2 m_{\cal P}^2 m_0^2 f_{\cal P} \dd
\Big\{\mathbb{A}(u_0) + 32 \Big[i_1({\cal A}_\parallel,1) +
i_1({\cal A}_\perp,1) + \beta i_1({\cal V}_\parallel,1) \nnb \\
\ar \beta i_1({\cal V}_\perp,1) - 2 i_1({\cal A}_\parallel,v) -
2 i_1({\cal A}_\perp,v) \Big] +
4 (1+2\beta) \Big[ 2 i_2({\cal A}_\perp,1) + 
i_2({\cal V}_\parallel,1)\Big] \nnb \\
\ar 4 (2+\beta) \Big[ i_2({\cal A}_\parallel,1) + 2 i_2({\cal V}_\perp,1)       
- 2 i_2({\cal A}_\parallel,v) - 4 i_2({\cal V}_\perp,v)\Big] -
4 \widetilde{i}_4({\Bbb{B}}) \Big\} \nnb \\
\ek {(1-\beta)\over 72 \sqrt{6} M^4} m_Q^3 m_0^2 \mu_{\cal P} \dd \Big[
3 i_2({\cal T},1) - (1-\widetilde{\mu}_{\cal P}^2)\phi_\sigma(u_0) \Big]
\nnb \\
%
%
\ek {1\over 3 \sqrt{6} M^2} m_Q^2 m_{\cal P}^2 f_{\cal P} \dd
\Big\{i_1({\cal A}_\parallel,1) + i_1({\cal A}_\perp,1) +
\beta \Big[i_1({\cal V}_\parallel,1) + i_1({\cal V}_\perp,1)\Big] \nnb \\
\ek 2 \Big[i_1({\cal A}_\parallel,v) + i_1({\cal A}_\perp,v)\Big] \Big\}
\nnb \\
\ar {1 \over 288 \sqrt{6} M^2} \Big\{12 (1-\beta) m_Q \mu_{\cal P} m_0^2 \dd
i_2({\cal T},1)      
+(2-\beta) f_{\cal P} m_{\cal P}^2 m_0^2 \dd \widetilde{i}_4({\Bbb{B}}) \nnb \\
\ek 12 (2+\beta) m_Q^2 f_{\cal P} \dd \Big[ m_{\cal P}^2 \mathbb{A}(u_0) +      
m_0^2 \phi_\eta(u_0) \Big] \Big\} \nnb \\
%
%
\ar {\beta\over 16 \sqrt{6} \pi^2} m_Q m_{\cal P}^4 f_{\cal P}
\Big\{ \Big[      
i_1({\cal A}_\parallel,1) + i_1({\cal A}_\perp,1) - 
i_1({\cal V}_\parallel,1) - i_1({\cal V}_\perp,1)\Big] (m_Q^2 I_2 - I_1)
\Big\} \nnb \\
\ar {1\over 12 \sqrt{6} } \dd \Big\{
2 (1-\beta) m_Q \mu_{\cal P} i_2({\cal T},1) +
(1+2 \beta) m_{\cal P}^2 f_{\cal P} \Big[2 i_2({\cal A}_\perp,1) +
i_2({\cal V}_\parallel,1) \Big] \nnb \\
\ar (2+\beta) m_{\cal P}^2 f_{\cal P} \Big[i_2({\cal A}_\parallel,1) +
2 i_2({\cal V}_\perp,1) -2 i_2({\cal A}_\parallel,v) - 
4 i_2({\cal V}_\perp,v) \Big] \Big\} \nnb \\
\ek {1\over 144 \sqrt{6} } \dd \Big\{6 m_{\cal P}^2 f_{\cal P} \Big[
(2+\beta) \mathbb{A}(u_0) + 2 \widetilde{i}_4({\Bbb{B}}) \Big]+
3 (3 + 2 \beta) m_0^2 f_{\cal P} \phi_\eta(u_0) \nnb \\
\ar 8 (1-\beta) m_Q \mu_{\cal P} (1-\widetilde{\mu}_{\cal P}^2) \phi_\sigma(u_0) 
\Big\}~,
\eea

where,
\bea
\label{nolabel}
i_1(\phi,f(v)) \es \int {\cal D}\alpha_i \int_0^1 dv
\phi(\alpha_{\bar{q}},\alpha_q,\alpha_g) f(v) \theta(k-u_0)~, \nnb \\
i_2(\phi,f(v)) \es \int {\cal D}\alpha_i \int_0^1 dv
\phi(\alpha_{\bar{q}},\alpha_q,\alpha_g) f(v) \delta(k-u_0)~, \nnb \\
i_3(\phi,f(v)) \es \int {\cal D}\alpha_i \int_0^1 dv
\phi(\alpha_{\bar{q}},\alpha_q,\alpha_g) f(v) \delta^\prime(k-u_0)~, \nnb \\
\widetilde{i}_4(f(u)) \es \int_{u_0}^1 du f(u)~, \nnb \\
I_n \es \int_{m_Q^2}^\infty ds \,
{e^{m_Q^2/M^2 - s/M^2}\over s^n}~,\nnb
\eea
and
\bea
k \es \alpha_q + \alpha_g \bar{v}~,~~~~~u_0={M_1^2 \over M_1^2
+M_2^2}~,~~~~~M^2={M_1^2 M_2^2 \over M_1^2
+M_2^2}~,\nnb \\ \nnb \\
\mu_{\cal P} \es f_{\cal P} {m_{\cal P}^2 \over m_{q_1} + m_{q_2}}~,~~~~~
\widetilde{\mu}_{\cal P} = {m_{q_1} + m_{q_2} \over m_{\cal P}}~,~~~~~
\eea
The  $D\alpha_i = d\alpha_{\bar q} d\alpha_q d\alpha_g
\delta(1-\alpha_{\bar q} - \alpha_q - \alpha_g)$, $q_1$ and $q_2$ are the light quarks, $Q$ is the heavy quark,  the subscript ${\cal P}$ stands for pseudoscalar meson 
and the functions  ${\cal A}_\parallel$, ${\cal A}_\perp$, 
${\cal T}$, ${\cal V}_\parallel$,  ${\cal V}_\perp$, $\phi_\sigma$,
$\phi_\sigma^\prime$, $\phi_\eta$, $\phi_{\cal P}$, $\mathbb{A}$ and
$\Bbb{B}$ are the DAs with definite twists for the pseudoscalar mesons.
To shorten the above expression, we have ignored the light quarks masses as well as terms containing gluon condensates, but we take into account their contribution when doing
numerical analysis. The continuum subtraction is performed using  results of the work \cite{Rh32v15}.

\section{Numerical analysis}

This section is devoted to the numerical analysis of the strong coupling
constants of  mesons with spin--3/2 and spin--1/2 heavy baryons.
The main input parameters for performing numerical analysis  are the DAs of the light pseudoscalar mesons, whose  expressions
are presented in \cite{Rh32v19}. The other input parameters
appearing in the sum rules are, $\qq=-(0.24\pm 0.001)^3~GeV^3$, $m_0^2 = (0.8
\pm 0.2)~GeV^2$ \cite{Rh32v13}, $f_\pi = 0.131~GeV$, $f_K=0.16~GeV$ and
$f_\eta=0.13~GeV$.

In the sum rules for the strong coupling constants of light pseudoscalar 
mesons with
heavy baryons, there are three auxiliary
 parameters, namely Borel mass $M^2$,
continuum threshold $s_0$ and the arbitrary parameter $\beta$ in the
expressions of the interpolating currents of spin--1/2 baryons.
It is clear that, any physical quantity, like the aforementioned
strong coupling constants, should be independent of these auxiliary parameters. Therefore, we try to
find so called ``working regions" of these  parameters, where
$g_{B^*BP}$ is practically independent of them. The upper limit of $M^2$ can be
obtained by demanding that the higher states and continuum contributions
contribute less than, say, 50\% of the total dispersion integral. The lower
bound of $M^2$ can be determined by requiring that the highest power in $1/M^2$
should be less than (20--25)\% of the highest power $M^2$. These two
conditions allow us to fix the following working regions: $15~GeV^2 \le M^2
\le 30~GeV^2$ for the bottom baryons, and $4~GeV^2 \le M^2 \le 12~GeV^2$
for the charmed baryons.      
As far as continuum threshold is concerned, we choose it in the interval between
$s_0=(m_B+0.5)^2~GeV^2$ and $s_0=(m_B+0.7)^2~GeV^2$.

\begin{figure}[t]
\begin{center}
\scalebox{0.5}{\includegraphics{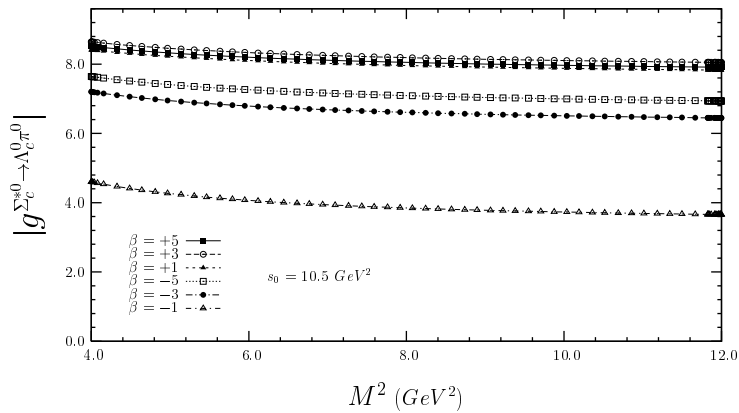}}
\end{center}
\caption{The dependence of the strong coupling constant for 
the $\Sigma_c^{*0} \rar \Lambda_c^{0} \pi^0$ transition on $M^2$ 
at several fixed values of $\beta$, and at 
$s_0=10.5~GeV^2$.}
\end{figure}

\begin{figure}[b]
\begin{center}
\scalebox{0.5}{\includegraphics{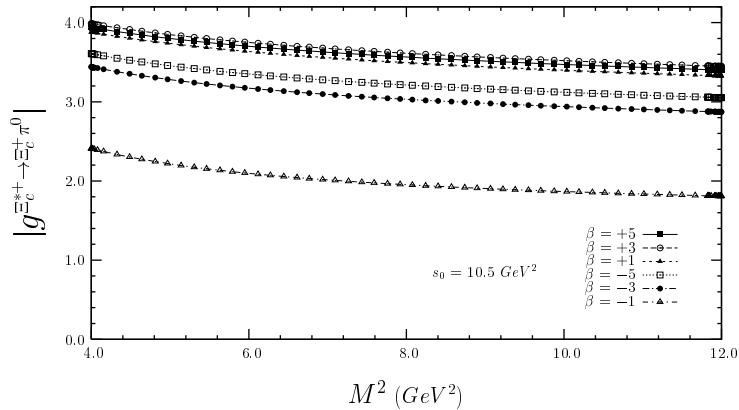}}
\end{center}
\caption{The dependence of the strong coupling constant for 
the $\Xi_c^{*+} \rar \Xi_c^{+} \pi^0$ transition on $M^2$ 
at several  fixed values of $\beta$, and at 
$s_0=10.5~GeV^2$.}
\end{figure}

\begin{figure}[t]
\begin{center}
\scalebox{0.5}{\includegraphics{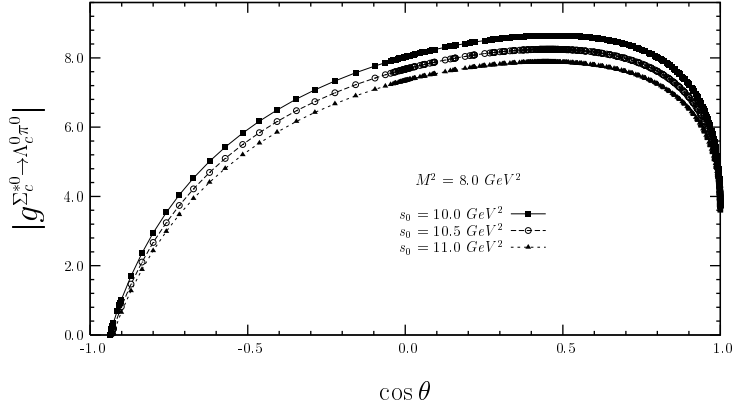}}
\end{center}
\caption{The dependence of the strong coupling constant for 
the $\Sigma_c^{*0} \rar \Lambda_c^{0} \pi^0$ transition on $\cos\theta$ 
at several  fixed values of $s_0$, and at $M^2= 8.0~GeV^2$.}
\end{figure}

\begin{figure}[b]
\begin{center}
\scalebox{0.5}{\includegraphics{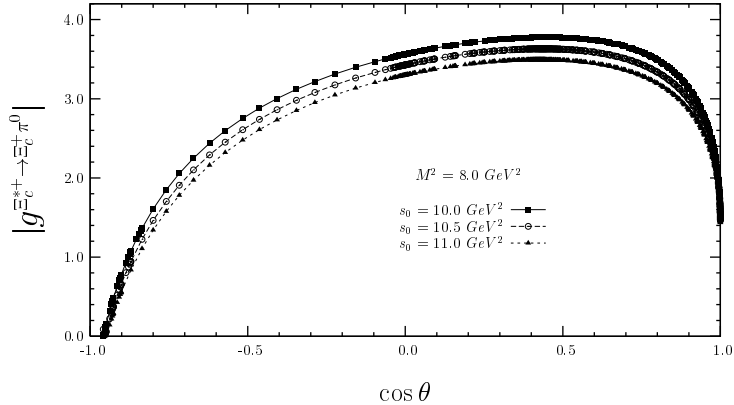}}
\end{center}
\caption{The dependence of the strong coupling constant for 
the $\Xi_c^{*+} \rar \Xi_c^{+} \pi^0$ transition on $\cos\theta$ 
at several  fixed values of $s_0$, and at $M^2= 8.0~GeV^2$.}
\end{figure}

As an example, in Figs. 1 and 2, we present the dependence of 
the coupling constants for the
$\Sigma_c^{*0} \to \Lambda_c^0 \pi^0$ and $\Xi_c^{*+} \to \Xi_c^+
\pi^0$ transitions on $M^2$, at different fixed values of  $\beta$ and at
$s_0=10.5~GeV^2$. From these figures, we see that the coupling constants 
for these transitions exhibit good stability when $M^2$ is varied in the above
mentioned ``working region''. 
Depicted in Figs. 2 and 4 are
the dependences of the same coupling constants on $\cos\theta$ at several
fixed values of $s_0$ and at $M^2=8~GeV^2$, where $\beta=\tan\theta$. From these figures, we
observe  that when $\cos\theta$ is varied in the domain
$-0.3 \le \cos\theta \le 0.5$, the coupling constants show rather stable
behavior  and they also have very weak dependence on $s_0$. Similar analysis for the strong coupling constants of all S$^*$SP and
S$^*$AP is performed and the results are presented in Tables 3 and 4, respectively.
For completeness, in these Tables we also present the predictions of the 
Ioffe current $(\beta=-1)$ for these coupling constants. Here, we would like
to remind that, our obtained domain for $\cos\theta$ lies inside the more wide interval obtained from  analysis of mass sum rules for heavy non strange
baryons in \cite{Rh32v18,Rh32v25}. 

Note that only few of the presented coupling constants can be measured
directly from the analysis of the decays, and the remaining coupling
constants can only be measured, indirectly.
At present, the decay widths of the $\Sigma_c^{*++} \to \Lambda_c^+ \pi^+$
and $\Sigma_c^{*0} \to \ \Lambda_c^+ \pi^-$ are measured, experimentally and
also the upper bounds for the $\Sigma_c^{*+} \to \Lambda_c^+ \pi^0$,
$\Xi_c^{*+} \to \Xi_c^{0} \pi^+$, $\Xi_c^{*+} \to \Xi_c^{+} \pi^0$,
$\Xi_c^{*0} \to \Xi_c^{+} \pi^-$ and $\Xi_c^{*0} \to \Xi_c^{0} \pi^0$ are
announced (see \cite{Rh32v04} and \cite{Rh32v06}).
Using the matrix element for the $3/2 \to 1/2 \pi$ transition, i.e.,
\bea
\label{nolabel}
{\cal M} = g_{B^*B\pi} \bar{u} (p) u^\alpha (p+q) q_\alpha~, 
\eea
one can easily obtain the following relation for the corresponding decay width:
\bea
\label{eh32v21}
\Gamma = {g^2\over 24 \pi m_1^2} \vel \vec{q} \ver^3 \Big[ (m_1+m_2)^2 -
m_\pi^2 \Big]~,
\eea
where $\vel\vec{q}\ver$ is the momentum of the $\pi$ meson. Using the values of the coupling
constants from Tables (3) and (4), and also Eq. (\ref{eh32v21}), we can easily
predict the values of the corresponding decay widths. Our
predictions on these decays, the experimental results, as well as
predictions of other approaches on these coupling constants are presented in
Table 5.
From this Table, we see that our predictions on decay widths for the
above--mentioned kinematically allowed transitions are all in good agreement
with the existing experimental results and the prediction of other approaches.        

In summary, we calculated the strong coupling constants of
spin--3/2 to spin--1/2 transitions with the participation of pseudoscalar
mesons within LCSR. Our analysis shows that all S$^*$SP and S$^*$AP
couplings are described by only one invariant function in each class of
transitions. Moreover, we estimated the widths of the kinematically allowed
transitions, which match quite good with  the existing
experimental data, as well as predictions of other approaches.


\begin{table}[tbh]

\renewcommand{\arraystretch}{1.3}
\addtolength{\arraycolsep}{-0.5pt}
\small
$$
\begin{array}{|l|r@{\pm}l|r@{\pm}l||l|r@{\pm}l|r@{\pm}l|}
\hline \hline  
 \multirow{2}{*}{$g^{\mbox{\small{\,channel}}}$}        &\multicolumn{4}{c||}{\mbox{Bottom Baryons}}   &  
 \multirow{2}{*}{$g^{\mbox{\small{\,channel}}}$}        &\multicolumn{4}{c|}{\mbox{Charmed Baryons}} \\
	                                                &   \multicolumn{2}{c}{\mbox{~General current~}}        & 
	                                                    \multicolumn{2}{c||}{\mbox{~Ioffe current~}}       & &
                                                                                                                   \multicolumn{2}{|c}{\mbox{~General current~}}  & 
                                                                                                                   \multicolumn{2}{c|}{\mbox{~Ioffe current~}}       \\ \hline
 g^{\Xi_b^{*0}    \rar \Xi_b^{'0}    \pi^0}     &~~~~~~~2.0&0.4    &~~~~~1.6&0.4 &
 g^{\Xi_c^{*+}    \rar \Xi_c^{'+}    \pi^0}     &~~~~~~ 2.1&0.3    &~~~~ 2.3&0.4   \\ 
 g^{\Sigma_b^{*0} \rar \Sigma_b^-    \pi^+}     &       3.7&0.6    &     3.5&0.5  &
 g^{\Sigma_c^{*+} \rar \Sigma_c^0    \pi^+}     &       4.2&0.5    &     4.3&0.4   \\
 g^{\Xi_b^{*0}    \rar \Sigma_b^+     K^-}      &       4.8&1.0    &     2.6&0.5  &
 g^{\Xi_c^{*+}    \rar \Sigma_c^{++}  K^-}      &       4.5&0.5    &     3.9&0.4   \\
 g^{\Omega_b^{*-} \rar \Xi_b^{'0}     K^-}      &       5.0&1.4    &     2.5&0.4  &
 g^{\Omega_c^{*0} \rar \Xi_c^{'+}     K^-}      &       4.3&0.5    &     4.5&0.4   \\
 g^{\Sigma_b^{*+} \rar \Sigma_b^+    \eta_1}    &       2.9&0.7    &     2.0&0.4  &
 g^{\Sigma_c^{*++}\rar \Sigma_c^{++} \eta_1}    &       2.9&0.4    &     2.7&0.4   \\
 g^{\Xi_b^{*0}    \rar \Xi_b^{'0}    \eta_1}    &       1.3&0.3    &     0.9&0.3  &
 g^{\Xi_c^{*+}    \rar \Xi_c^{'+}    \eta_1}    &       1.5&0.3    &     1.0&0.2   \\
 g^{\Omega_b^{*-} \rar \Omega_b^-    \eta_1}    &       5.8&1.6    &     3.6&0.6  &
 g^{\Omega_c^{*0} \rar \Omega_c^0    \eta_1}    &       5.9&0.8    &     5.8&0.4   \\
 \hline \hline
\end{array}
$$

\caption{Values of the strong coupling constants $g$ in $GeV^{-1}$ for the transitions
among the sextet spin--3/2 and sextet spin--1/2 heavy baryons with pseudoscalar mesons.}

\renewcommand{\arraystretch}{1}
\addtolength{\arraycolsep}{-1.0pt}

\end{table}


\begin{table}[tbh]

\renewcommand{\arraystretch}{1.3}
\addtolength{\arraycolsep}{-0.5pt}
\small
$$
\begin{array}{|l|r@{\pm}l|r@{\pm}l||l|r@{\pm}l|r@{\pm}l|}
\hline \hline  
 \multirow{2}{*}{$g^{\mbox{\small{\,channel}}}$}        &\multicolumn{4}{c||}{\mbox{Bottom Baryons}}   &  
 \multirow{2}{*}{$g^{\mbox{\small{\,channel}}}$}        &\multicolumn{4}{c|}{\mbox{Charmed Baryons}} \\
	                                                &   \multicolumn{2}{c}{\mbox{~General current~}}        & 
	                                                    \multicolumn{2}{c||}{\mbox{~Ioffe current~}}       & &
                                                                                                                   \multicolumn{2}{|c}{\mbox{~General current~}}  & 
                                                                                                                   \multicolumn{2}{c|}{\mbox{~Ioffe current~}}       \\ \hline
 g^{\Xi_b^{*0}    \rar \Xi_b^0       \pi^0}     &~~~~~~~3.0&0.6    &~~~~~1.4&0.3  &
 g^{\Xi_c^{*+}    \rar \Xi_c^+       \pi^0}     &~~~~~~ 3.5&0.5    &~~~~ 2.0&0.3   \\
 g^{\Sigma_b^{*-} \rar \Lambda_b^0   \pi^-}     &       6.0&1.1    &     2.5&0.5  &
 g^{\Sigma_c^{*0} \rar \Lambda_c^+   \pi^-}     &       7.8&1.0    &     3.9&0.6   \\
 g^{\Sigma_b^{*0} \rar \Xi_b^0       \bar{K}^0} &       3.7&0.5    &     2.0&0.5  &
 g^{\Sigma_c^{*+} \rar \Xi_c^+       \bar{K}^0} &       5.0&1.0    &     3.1&0.4   \\
 g^{\Omega_b^{*-} \rar \Xi_b^-       \bar{K}^0} &       5.0&0.8    &     2.6&0.4  &
 g^{\Omega_c^{*0} \rar \Xi_c^0       \bar{K}^0} &       6.2&1.5    &     4.1&0.5   \\
 g^{\Xi_b^{*0}    \rar \Xi_b^-        K^+}      &       3.6&0.5    &     1.9&0.6  &
 g^{\Xi_c^{*+}    \rar \Xi_c^0        K^+}      &       4.4&0.8    &     3.0&0.4   \\
 g^{\Xi_b^{*0}    \rar \Xi_b^0       \eta_1}    &       5.4&1.0    &     2.5&0.4  &
 g^{\Xi_c^{*+}    \rar \Xi_c^+       \eta_1}    &       6.9&1.5    &     4.0&0.5   \\
 \hline \hline
\end{array}
$$

\caption{Values of the strong coupling constants $g$ in $GeV^{-1}$ for the transitions
among the sextet spin--3/2 and anti--triplet spin--1/2 heavy baryons with pseudoscalar mesons.}

\renewcommand{\arraystretch}{1}
\addtolength{\arraycolsep}{-1.0pt}

\end{table}


\begin{table}[tbh]

\renewcommand{\arraystretch}{1.3}
\addtolength{\arraycolsep}{-0.5pt}
\small
$$
\begin{array}{|l|c|c|c|c|c|c|}
\hline \hline  
                    &\mbox{\rm Our work}
                    &\mbox{\rm CQM \cite{Rh32v26}}
                    &\mbox{\rm LFQM \cite{Rh32v27}}
                    &\mbox{\rm RQM \cite{Rh32v28}}
                    &\mbox{\rm NRQM \cite{Rh32v29}}
                    &\mbox{\rm Experiment\cite{Rh32v30}} \\ \hline
 \Gamma(\Sigma_c^{*++} \to \Lambda_c^+ \pi^+) & 14.6 \pm3.8 & 20       & 12.84     & 21.90 \pm 0.87 & 17.52 \pm 0.74 & 14.9 \pm 1.9 \\ 
 \Gamma(\Sigma_c^{*+} \to \Lambda_c^+ \pi^0)  & 14.6 \pm3.8 & 20       & \mbox{--} & \mbox{--}      & \mbox{--}      &   < 17       \\
 \Gamma(\Sigma_c^{*0} \to \Lambda_c^0 \pi^0)  & 14.6\pm3.8  & 20       & 12.40     & 21.20 \pm 0.81 & 16.90 \pm 0.71 & 16.1 \pm 2.1 \\
 \Gamma(\Xi_c^{*+} \to \Xi_c^0 \pi^+)         &  2.8 \pm0.9 & \mbox{--}& 1.12      &  1.78 \pm 0.33 & \mbox{--}      &   < 3.1      \\
 \Gamma(\Xi_c^{*+} \to \Xi_c^+ \pi^0)         &  1.4 \pm 0.4& \mbox{--}& 0.69      &  1.26 \pm 0.17 & \mbox{--}      &   < 3.1      \\
 \Gamma(\Xi_c^{*0} \to \Xi_c^+ \pi^-)         &  2.8 \pm0.9 & \mbox{--}& 1.16      &  2.11 \pm 0.29 & \mbox{--}      &   < 5.5      \\
 \Gamma(\Xi_c^{*0} \to \Xi_c^0 \pi^0)         &  1.4\pm0.4  & \mbox{--}& 0.72      &  1.01 \pm 0.15 & \mbox{--}      &   < 5.5      \\
\hline \hline
\end{array}
$$
\caption{Strong one--pion decay rates. Here the short keys stand for:
(CQM) Constituent Quark Model, (LFQM) Light--Front Quark Model,
(RQM) Relativistic Quark Model, (NRQM) Non--Relativistic Quark Model.
The results are presented in units of $MeV$.}
\renewcommand{\arraystretch}{1}
\addtolength{\arraycolsep}{-1.0pt}

\end{table}


\newpage
\bAPP{A}{}

In this appendix we present the expressions of the correlation functions
in terms of invariant function $\Pi_1^{(1)}$ and $\Pi_1^{(2)}$ involving 
$\pi$, $K$ and  $\eta_1$ mesons.

\begin{itemize}
\item Correlation functions describing pseudoscalar mesons with
sextet--sextet baryons.
\end{itemize}
\baeeq
\label{nolabel}
\Pi^{\Sigma_b^{*+} \rar \Sigma_b^+ \pi^0 } \es
\sqrt{2} \Pi_1^{(1)}(u,u,b)~, \nnb \\
\Pi^{\Sigma_b^{*-} \rar \Sigma_b^- \pi^0 } \es
- \sqrt{2} \Pi_1^{(1)}(d,d,b)~, \nnb \\
\Pi^{\Sigma_b^{*+} \rar \Sigma_b^0 \pi^+ } \es
\sqrt{2} \Pi_1^{(1)}(d,u,b)~, \nnb \\
\Pi^{\Sigma_b^{*0} \rar \Sigma_b^- \pi^+ } \es
\sqrt{2} \Pi_1^{(1)}(u,d,b)~, \nnb \\
\Pi^{\Xi_b^{*0} \rar \Xi_b^{'-}     \pi^+ } \es
\Pi_1^{(1)}(d,s,b)~, \nnb \\
\Pi^{\Sigma_b^{*0} \rar \Sigma_b^+ \pi^- } \es
\sqrt{2} \Pi_1^{(1)}(d,u,b)~, \nnb \\
\Pi^{\Sigma_b^{*-} \rar \Sigma_b^0 \pi^- } \es
\sqrt{2} \Pi_1^{(1)}(u,d,b)~, \nnb \\
\Pi^{\Xi_b^{*-} \rar \Xi_b^{'0}     \pi^- } \es
\Pi_1^{(1)}(u,s,b)~, \nnb \\
\Pi^{\Xi_b^{*0} \rar \Sigma_b^+ K^- } \es
\sqrt{2} \Pi_1^{(1)}(u,u,b)~, \nnb \\
\Pi^{\Xi_b^{*-} \rar \Sigma_b^0 K^- } \es
\sqrt{2} \Pi_1^{(1)}(u,d,b)~, \nnb \\
\Pi^{\Omega_b^{*-} \rar \Xi_b^{'0}     K^- } \es
\sqrt{2} \Pi_1^{(1)}(s,s,b)~, \nnb \\
\Pi^{\Sigma_b^{*+} \rar \Xi_b^{'0}  K^+  } \es
\sqrt{2} \Pi_1^{(1)}(u,u,b)~, \nnb \\
\Pi^{\Sigma_b^{*0} \rar \Xi_b^{'-}  K^+  } \es
\Pi_1^{(1)}(u,d,b)~, \nnb \\
\Pi^{\Xi_b^{*0} \rar \Omega_b^{-} K^+  } \es
\sqrt{2}  \Pi_1^{(1)}(s,s,b)~, \nnb \\
\Pi^{\Xi_b^{*0} \rar \Sigma_b^0 \bar{K}^0  } \es
\Pi_1^{(1)}(d,u,b)~, \nnb \\
\Pi^{\Xi_b^{*-} \rar \Sigma_b^- \bar{K}^0  } \es
\sqrt{2} \Pi_1^{(1)}(d,d,b)~, \nnb \\
\Pi^{\Omega_b^{*-} \rar \Xi_b^{'-}     \bar{K}^0 } \es
\sqrt{2} \Pi_1^{(1)}(s,s,b)~, \nnb \\
\Pi^{\Sigma_b^{*0} \rar \Xi_b^{'0}  K^0  } \es
\Pi_1^{(1)}(d,u,b)~, \nnb \\
\Pi^{\Sigma_b^{*-} \rar \Xi_b^{'-}  K^0  } \es
\sqrt{2} \Pi_1^{(1)}(d,d,b)~, \nnb \\
\Pi^{\Xi_b^{*-} \rar \Omega_b^{-} K^0  } \es
\sqrt{2}  \Pi_1^{(1)}(s,s,b)~, \nnb \\
\Pi^{\Sigma_b^{*+} \rar  \Sigma_b^+ \eta_1 } \es
{2\over \sqrt{6}}  \Pi_1^{(1)}(u,u,b)~, \nnb \\
\Pi^{\Sigma_b^{*-} \rar  \Sigma_b^- \eta_1 } \es
{2\over \sqrt{6}}  \Pi_1^{(1)}(d,d,b)~, \nnb \\
\Pi^{\Xi_b^{*-} \rar \Xi_b^{'-}  \eta_1  } \es
{1\over \sqrt{6}}  \Big[\Pi_1^{(1)}(d,s,b) - 2 \Pi_1^{(1)}(s,d,b) \Big]~, \nnb \\
\Pi^{\Omega_b^{*-} \rar \Omega_b^{-}      \eta_1 } \es
- {4 \over \sqrt{6}} \Pi_1^{(1)}(s,s,b)~.
\eaeeq

\begin{itemize}
\item Correlation functions responsible for the transitions of the
sextet--anti--triplet baryons.
\end{itemize}
\baeeq
\label{nolabel}
\Pi^{\Xi_b^{*-} \rar \Xi_b^0        \pi^- } \es
\Pi_1^{(2)}(d,s,b)~, \nnb \\
\Pi^{\Xi_b^{*0} \rar \Xi_b^-        \pi^+ } \es
\Pi_1^{(2)}(u,s,b)~, \nnb \\
\Pi^{\Sigma_b^{*0} \rar \Xi_b^0    \bar{K}^0 } \es
- \Pi_1^{(2)}(d,u,b)~, \nnb \\
\Pi^{\Sigma_b^{*-} \rar  \Xi_b^-    \bar{K}^0 } \es
- \Pi_1^{(2)}(d,d,b)~, \nnb \\
\Pi^{\Omega_b^{*-} \rar  \Xi_b^-    \bar{K}^0 } \es
\sqrt{2} \Pi_1^{(2)}(s,s,b)~, \nnb \\
\Pi^{\Xi_b^{*0} \rar   \Lambda_b^0   \bar{K}^0 } \es
- \Pi_1^{(2)}(d,u,b)~, \nnb \\
\Pi^{\Sigma_b^{*0} \rar  \Xi_b^0    K^0 } \es
- \Pi_1^{(2)}(d,u,b)~, \nnb \\
\Pi^{\Sigma_b^{*-} \rar  \Xi_b^-    K^0 } \es
- \sqrt{2} \Pi_1^{(2)}(d,d,b)~, \nnb \\
\Pi^{\Omega_b^{*-} \rar  \Xi_b^-    K^0 } \es
\sqrt{2} \Pi_1^{(2)}(s,s,b)~, \nnb \\
\Pi^{\Xi_b^{*0} \rar   \Lambda_b^0   K^0 } \es
- \Pi_1^{(2)}(d,u,b)~, \nnb \\
\Pi^{\Sigma_b^{*+} \rar   \Lambda_b^0   K^+ } \es
- \sqrt{2} \Pi_1^{(2)}(u,u,b)~, \nnb \\
\Pi^{\Sigma_b^{*0} \rar   \Xi_b^-    K^+ } \es
- \Pi_1^{(2)}(u,d,b)~, \nnb \\
\Pi^{\Xi_b^{*0} \rar   \Xi_b^-    K^+ } \es
\Pi_1^{(2)}(d,s,b)~, \nnb \\
\Pi^{\Sigma_b^{*-} \rar    \Lambda_b^0   K^- } \es
\sqrt{2} \Pi_1^{(2)}(d,d,b)~, \nnb \\
\Pi^{\Omega_b^{*-} \rar  \Xi_b^0    K^- } \es
\sqrt{2} \Pi_1^{(2)}(s,s,b)~, \nnb \\
\Pi^{\Xi_b^{*-} \rar   \Xi_b^0    K^- } \es
\Pi_1^{(2)}(u,s,b)~, \nnb \\
\Pi^{\Xi_b^{*-} \rar   \Xi_b^-    \eta_1 } \es
{1\over \sqrt{6}} \Big[\Pi_1^{(2)}(d,s,b) + 2 \Pi_1^{(2)}(s,d,b) \Big]~, \nnb \\
\eaeeq 

In the case of charmed baryons it is enough to make the replacement $b \rar
c$ and increas the charge of each baryon by a positive unit.

\eAPP

\end{document}